P. N. LEBEDEV PHYSICAL INSTITUTE OF
THE RUSSIAN ACADEMY OF SCIENCES

PREPRINT

**O.D. Dalkarov, A.S. Rusetskii,**

**S.A. Pikuz, T.A. Shelkovenko and I.N. Tilikin**

# Observation of neutron emission in the process of X-pinch

Moscow 2015

# Observation of neutron emission in the process of X-pinch

O.D. Dalkarov, A.S. Rusetskii*,

S.A. Pikuz, T.A. Shelkovenko and I.N. Tilikin

*Lebedev Physical Institute RAS, Moscow, Russia*

*\* Corresponding author. E-mail: rusets@lebedev.ru*

**Annotation**

The results of measuring the neutron flux in the process of the X-pinch are presented. The measurements were carried out using CR-39 track detectors. It was found that in the process of X-pinch recorded neutron emission over a wide energy range (from thermal to energies greater than 10 MeV) with an intensity of more than $10^8$ neutrons per shot into $4\pi$ sr solid angle (assuming isotropy of radiation and localizing the source in the "hot spot"). Data of track detectors suggest that at the time of discharge produced fast neutrons are then slowed down and turned into thermal.

**Introduction**

Earlier it was reported about registration at the facility ERG (Electronic Relativistic Generator) (maximum voltage of 1 MW, maximum current of about 10 kA at pulse duration of about 1 microsecond) bursts of fast neutrons in the process of high-voltage discharge in the air. [1] To register neutron fluxes we used CR-39 track detectors to assess the neutron flux in different energy ranges. It was found that the average flux of fast neutrons reaches values of $\sim 10^5$ / cm$^2$ / shot into $4\pi$ sr solid angle. Moreover, the neutron energy exceeds 10 MeV. The nature of neutron emission is not completely clear and the mechanism of their generation does not fit into existing concepts of atmospheric discharge. It has been suggested that the neutron emission is related to the acceleration processes occurring in the interaction of the plasma discharge from the electrode surface. To test the possible mechanisms of generation of nuclear radiation in a dense plasma neutron fluxes were measured on the high-current pulse generator BIN [2] (the maximum voltage of 300 kV, maximum current of 250 kA at a pulse width of 100 ns) with Mo Hybrid X-pinch load [3]. In this case, the power density reaches a value which is orders of magnitude larger than for a high voltage discharge installation ERG. At the same time the possible influence of impurities (deuterium, tritium, etc.) on the possibility of generating neutrons completely ruled by long vacuum pumping of the discharge chamber to the pressure 2 10$^{-4}$ torr.

**Experimental technique**

For the detection of neutrons used track detectors CR-39, which are insensitive to electromagnetic interference and low sensitive to X-rays and gamma radiation. This advantage makes it possible to use them in the process of high-voltage discharges and detectors placed at a minimum distance from the object (in this case from the "hot spot"). For additional protection against X-ray and gamma radiation track detectors were placed in a lead container. Arrangement of CR-39 track detectors on an X-pinch facility is shown in Fig. 1. Background detectors located around 5 m from the "hot spot".

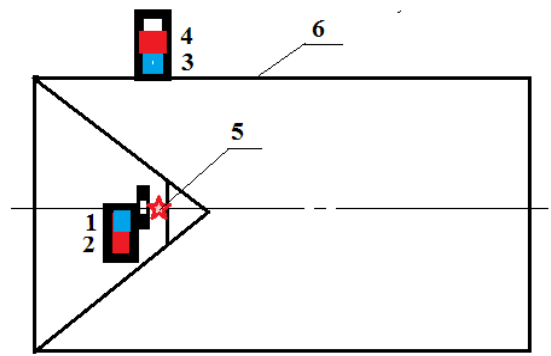

Fig.1. Arrangement of CR-39 track detectors on an X-pinch facility. 1,2,3,4 - track detectors (radiators: PE - 1 and 2, Na tetraborate solution in glycerol - 3 and 4), 5 - a "hot spot" X-pinch, 6 - outline of vacuum chamber.

For neutron detection method was used, described in detail in [1]. Track detectors are calibrated using a proton beam accelerator Van de Graaff ($E_p$ = 0.5 - 3.0 MeV), α-sources ($E_α$ = 2 - 7.7 MeV) and cyclotron beam ($E_α$ = 8 - 30 MeV) in the at Skobeltsyn Institute of Nuclear Physics of Moscow State University. After irradiation, the detectors were etched in a solution of 6M NaOH in $H_2O$ at 70°C for 7 hours. Detailed procedures for the calibration of charged particles is described in [4].

Calibrating the detector CR-39 with fast neutrons from a source Cf-252 with activity $3 \times 10^4$ n/s in the solid angle 4π sr showed that the average detection efficiency of fast neutrons track detector c PE radiator 120 μm was found to be $η_n$ = 5.7 x10$^{-5}$. The distribution of the diameters of the tracks of recoil protons after irradiation track detector CR-39 with neutrons is shown in Fig. 2. Tracks recoil protons have diameters of 4 - 8 μm. Calibration of neutron detector 14 MeV is described in detail in [5].

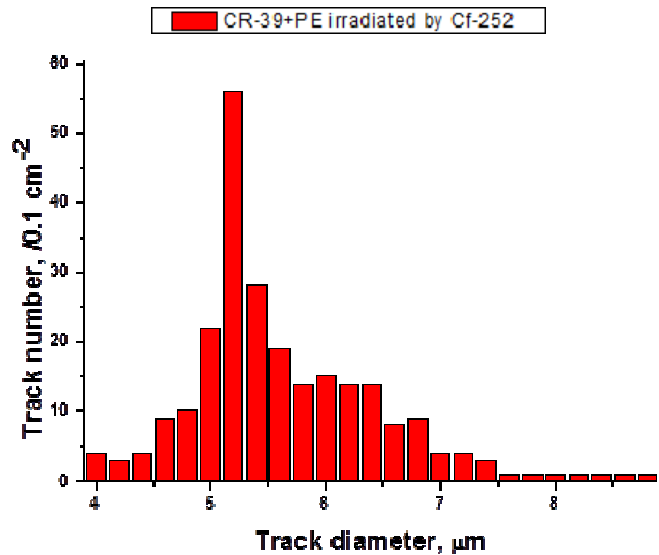

Fig.2. The distribution of the diameters of the recoil proton tracks after irradiation track detector CR-39 by neutrons from $^{252}$Cf source (7h etching detector 6M NaOH, at t = 70 ° C).

The main reactions taking place in the substance CR-39 track detector by neutrons of 14 MeV were described in [6]. To evaluate the neutron flux in different energy ranges, we used the following reactions (see. Table 1)

**Experimental results**

It was produced 3 "shots" in X-pinch installation. Then track detectors were etched under standard conditions and examined on an optical microscope MBI-9 with a CCD camera.

For detectors exposed near the facility, the number of single tracks significantly (3 - 6 times) exceeded the background values for the detectors located around 5 m from the X-pinch. Diameter distribution detectors of tracks on CR-39 in a vacuum chamber and the chamber lid after three shots shown in Fig. 3.

Table 1

| The type of reaction | Cross section of the reaction, mbarn | Efficient threshold, MeV | Radiator | Fast neutron detection efficiency of CR-39 track detector | Signature of reaction |
|---|---|---|---|---|---|
| $^{12}C(n,n')3\alpha$ | 202 ± 30 | 9.6 | PE | $1.2 \cdot 10^{-6}$ | Three tracks coming from one point |
| $^{12}C(n,\alpha)^9Be$ | 62 ± 15 | 5.8 | PE | $1.57 \cdot 10^{-6}$ * | Two tracks coming from one point |
| $^{16}O(n,\alpha)^{13}C$ | ~150 | 3.1 | PE | $1.57 \cdot 10^{-6}$ * | Two tracks coming from one point |
| $^1H(n,n')^1H$ | ~750 | Elastic scattering | PE | $5.7 \cdot 10^{-5}$ | Single track of proton |
| $^{10}B(n,^7Li)\alpha$ | 700 | Thermal neutrons | Na tetraborate solution in glycerol | $1.4 \cdot 10^{-6}$ | Single track of α-particle with an energy of <2 MeV |

*- for reactions $^{12}C(n, \alpha)^9Be$ and $^{16}O(n, \alpha)^{13}C$ decay signature look the same (two tracks emanating from a point), and therefore the neutron detection efficiency $1.57 \times 10^{-6}$ refers to the sum of the cross sections of the two reactions

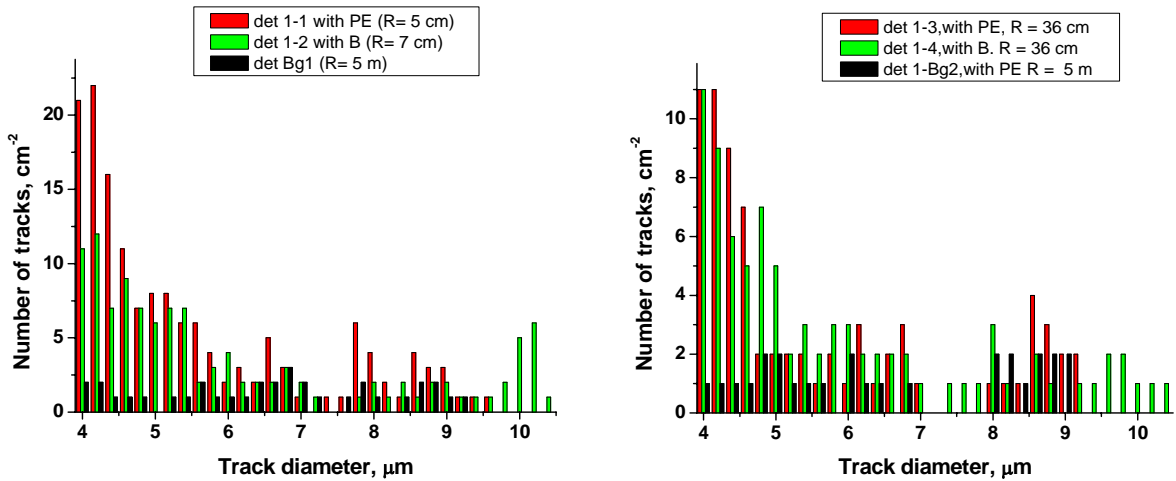

Fig. 3. Diameter distribution of tracks on CR-39 detectors in a vacuum chamber (left side) and the chamber lid (right) after 3 rounds. The detector with a radiator 120 μm polyethylene (dark bars). Indications of detectors with Na tetraborate in solution in glycerol (light bars). Background detectors were located 5 m from the "hot spot" (black bars). To construct the distribution we selected single tracks close to a circular shape corresponding to the direction close to normal incidence of particles.

Also, a significant excess above the background for the events received three and two tracks coming from one point. Representative photomicrographs of such events for the detectors located close to the "hot spot" shown in Fig. 4. These decays indicate a threshold reactions going by fast neutrons (see. Table 1). This gives an indication of the emission of neutrons with energies greater than 10 MeV. According to the track detectors evaluated neutron fluxes in various energy ranges (see. Table 2)

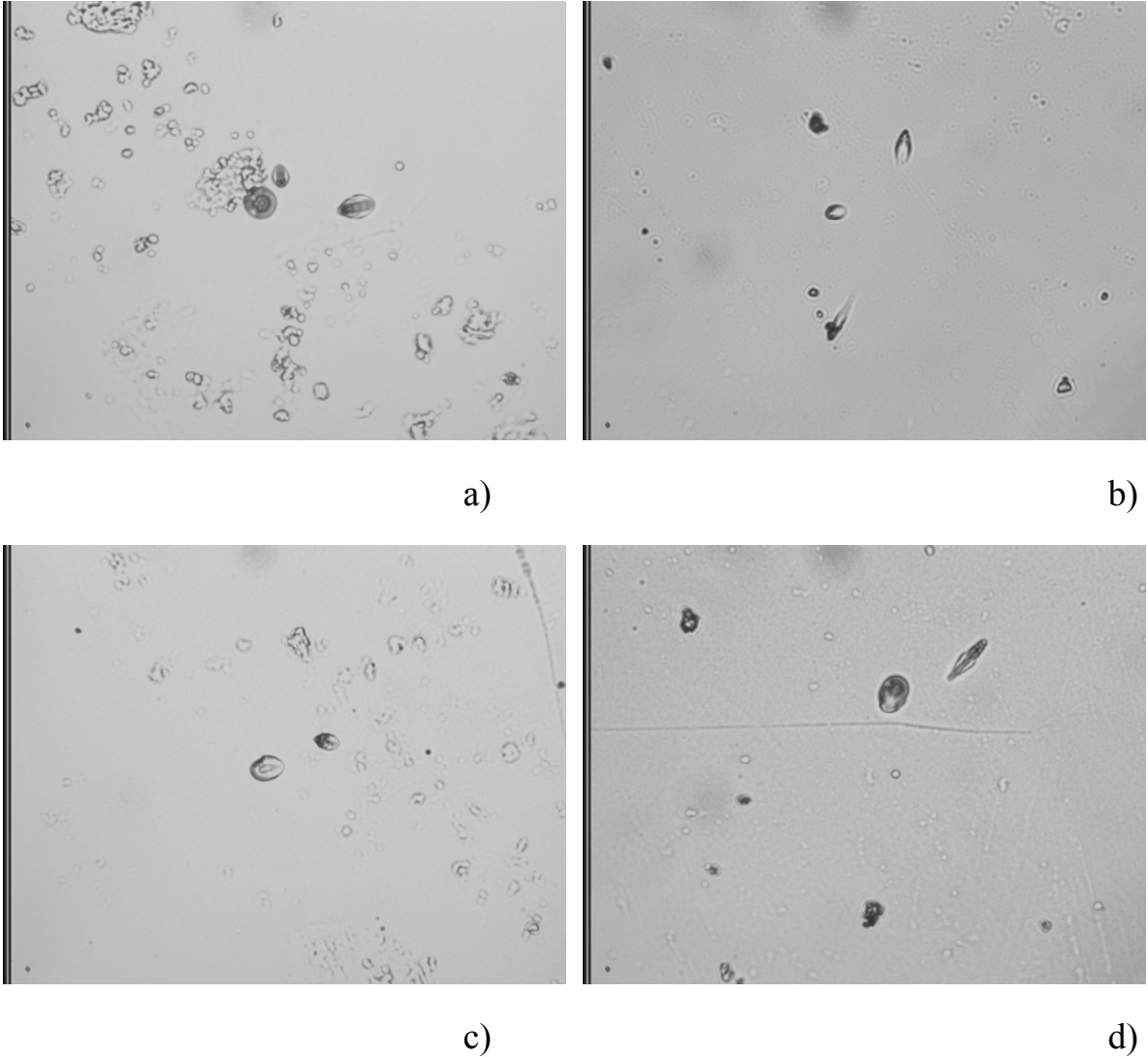

a)　　　　　　　　　　　　　　　　　　　　　　b)

c)　　　　　　　　　　　　　　　　　　　　　　d)

Fig. 4. The micrographs of events with three (a, b) and two (c, d) tracks coming from a single point. The image size of 120 x 100 microns.

**Conclusions**

1) Measurement of track detector CR-39 for the first time found that in the process of X-pinch recorded neutron emission over a wide energy range (from thermal to energies greater than 10 MeV) with an intensity of more than $10^8$ neutrons per shot into $4\pi$ sr solid angle (assuming isotropic radiation and localizing the source in the "hot spot").

2) Data detector in the vacuum chamber and its yield is an indication that the source of neutrons has strong anisotropy in the emission of neutrons. It is also possible that the additional emission of neutrons coming from the walls of the chamber under the influence of powerful X-ray radiation.

3) The data track detectors suggest that at the time of discharge produced fast neutrons that are then slowed down and turned into thermal.

4) To elucidate the mechanism of the observed neutron emission and clarify their location requires additional experiments.

Table 2. Fluxes of neutrons of different energies in the field of detectors

| Index of detector | Radiator | Effective distance from "hot spot", cm | $E_n > 1$ MeV 1/cm$^2$/shot | $E_n > 6$ MeV 1/cm$^2$/shot | $E_n > 10$ MeV 1/cm$^2$/shot | $E_n \sim 0/025$ eV 1/cm$^2$/shot |
|---|---|---|---|---|---|---|
| 1-1 | PE | 5.0 | 7.4 10$^5$ | 4.2 10$^5$ | 3.7 10$^5$ | - |
| 1-2 | A solution of sodium tetraborate in glycerol | 7.0 | 3.8 10$^5$ | 4.2 10$^5$ | 4.8 10$^5$ | 4.5 10$^6$ |
| 1-3 | PE | 36 | 2.9 10$^5$ | 3.2 10$^5$ | 2.4 10$^5$ | - |
| 1-4 | A solution of sodium tetraborate in glycerol | 36 | 3 10$^5$ | 3.2 10$^5$ | - | 1.7 10$^6$ |